\documentclass[12pt,preprint,a4paper]{aastex}

\usepackage{graphics,epsf}
\usepackage{amsmath}                
\usepackage{amsfonts}               
\usepackage{amssymb}                
\usepackage{epsfig}                 

\def \cm{~\rm{cm}}
\def \s{~\rm{s}}
\def \km{~\rm{km}}

\def \K{~\rm{K}}

\def \AU{~\rm{AU}}
\def \erg{~\rm{erg}}

\def \yr{~\rm{yr}}

\def \days{~\rm{day}}

\def \keV{~\rm{keV}}
\def \astrobj#1{#1}

\begin{document}

\title{USING X-RAY OBSERVATIONS TO EXPLORE THE BINARY INTERACTION IN ETA CARINAE}

\author{Amit Kashi\altaffilmark{1} and Noam Soker\altaffilmark{1}}

\altaffiltext{1}{Department of Physics, Technion$-$Israel Institute of
Technology, Haifa 32000 Israel; kashia@physics.technion.ac.il;
soker@physics.technion.ac.il.}
\setlength{\columnsep}{1cm}
\small

\begin{abstract}
We study the usage of the X-ray light curve, column density toward the hard X-ray source,
and emission measure (density square times volume),
of the massive binary system $\eta$ Carinae to determine the orientation of its semi-major axis.
The source of the hard X-ray emission is the shocked secondary wind.
We argue that, by itself, the observed X-ray flux cannot teach us much about the
orientation of the semi-major axis.
Minor adjustment of some unknown parameters of the binary system allows to fit the
X-ray light curve with almost any inclination angle and orientation.
The column density and X-ray emission measure, on the other hand, impose strong
constrains on the orientation.
We improve our previous calculations and show that the column density is more
compatible with an orientation where for most of the time the
secondary$-$the hotter, less massive star$-$is behind the primary star.
The secondary comes closer to the observer only for a short time near
periastron passage.
The ten-week X-ray deep minimum, which results from a large decrease in the emission measure,
implies that the regular secondary wind is substantially suppressed during that period.
This suppression is most likely resulted by accretion of mass from
the dense wind of the primary luminous blue variable (LBV) star.
The accretion from the equatorial plane might lead to the formation of a polar outflow.
We suggest that the polar outflow contributes to the soft X-ray emission during
the X-ray minimum; the other source is the shocked secondary wind in the tail.
The conclusion that accretion occurs at each periastron passage, every five and
a half years, implies that accretion had occurred at a much higher rate during the
Great Eruption of $\eta$ Car in the 19th century.
This has far reaching implications for major eruptions of LBV stars.
\end{abstract}


\section{INTRODUCTION}
\label{sec:intro}

The $P=5.54 \yr$ ($P =2022.7 \pm 1.3~$d; Damineli et al. 2008a)
periodicity of the massive binary system \astrobj{$\eta$ Car} is
observed in the entire electromagnetic band
(e.g., radio, Duncan \& White 2003; IR, Whitelock et al. 2004;
visible, van Genderen et al. 2006, Fernandez Lajus et al. 2008;
emission and absorption lines, Damineli et al. 1997, 2008a, b;
X-ray, Corcoran 2005).
The periodicity follows the 5.54~years periodic change in the orbital
separation in this highly eccentric, $e \simeq 0.9$, binary system
(e.g., Hillier et al. 2006).

It is generally agreed that the orbital plane lies in the
equatorial plane of the bipolar structure$-$the Homunculus, such
that the inclination angle (the angle between a line perpendicular
to the orbital plane and the line of sight) is $i=42$ (Davidson et
al. 2001; Smith 2002). However, there is a disagreement about the
orientation of the semimajor axis in the orbital plane$-$the
periastron longitude. We will use the commonly used periastron
longitude angle $\omega$: $\omega=0^\circ$ for a case when the
secondary is toward the observer at an orbital angle of $90^\circ$
before periastron, and so on, as summarized in equation
(\ref{eq:omega}),
\begin{equation}
\label{eq:omega}
\omega =
  \begin{cases}
~~0^\circ & \qquad  {\rm secondary~toward~observer~}90^\circ~{\rm before~periastron}
    \\
~90^\circ & \qquad  {\rm secondary~toward~observer~at~periastron}
    \\
180^\circ & \qquad  {\rm secondary~toward~observer~}90^\circ~{\rm after~periastron}
    \\
270^\circ & \qquad  {\rm secondary~toward~observer~at~apastron}.
 \end{cases}
\end{equation}

While some groups argue that the secondary (less massive) star is away
from us during periastron passages, $\omega=270^\circ$ (e.g., Nielsen
et al. 2007; Damineli et al. 2008), others argue that the secondary is
toward us during periastron passages, $\omega=90^\circ$
(Falceta-Gon\c{c}alves et al. 2005; Abraham et al. 2005; Kashi \& Soker 2007,
who use the angle $\gamma=90^\circ-\omega$).
Other semimajor axis orientations have also been proposed (Davidson
1997; Smith et al. 2004; Dorland 2007; Henley et al. 2008; Okazaki et al. 2008).

In a recent paper Kashi \& Soker (2008; hereafter KS08) examined a variety
of observations that shed light on the orientation of the semi-major axis:
(1) The Doppler shifts of some He~I P-Cygni lines that
are attributed to the secondary wind, of one Fe~II line that is attributed
to the primary wind, and of the Paschen emission lines that are attributed
to the shocked primary wind.
(2) The hydrogen column density toward the binary system as deduced from
X-ray observations by Hamaguchi et al. (2007; hereafter H07).
(3) The ionization of surrounding gas blobs by the radiation of the hotter secondary star.
KS08 found that all of these support an orientation where for most of the time the
secondary$-$the hotter less massive star$-$is behind the primary star ($\omega=90^\circ$).
The secondary comes closer to the observer only for a short time near
periastron passage.

In a more recent paper, Parkin et al. (2009, hereafter P09) built a model to fit the
X-ray cyclical light curve in the $2-10 \keV$ band as observed by RXTE (Corcoran 2005).
{} From their modelling they deduced that the secondary is away from us at, or somewhat after,
periastron ($\omega=270-300^\circ$).
This is an opposite orientation to that deduced by us in a previous paper (KS08).
In the present paper we take the challenge to compare the contradicting conclusions
of P09 and KS08.
For that we critically follow the arguments of P09 (sections \ref{sec:xray} and \ref{sec:collapse}),
and recheck and improve some of our previous calculations (section \ref{sec:N_H}).
We find severe problems with many of their assumptions.
We conclude that their model fails to account for the
X-ray observations.
We also introduce new calculations, and suggest that a polar outflow is responsible to some of the
observed soft x-ray emission (section \ref{sec:diss}) during the X-ray minimum.

\section{THE X-RAY EMISSION }
\label{sec:xray}

The hard X-ray emission observed in $\eta$ Car is emitted by the shocked secondary wind
(Corcoran et al. 2001; Pittard \& Corcoran 2002; Akashi et al. 2006, hereafter A06).
For constant winds' properties, the volume of the shocked secondary wind changes along
the orbit as $V \propto r^3$, where $r$ is the orbital separation.
The time scale for the shocked gas to flow out of the emitting volume goes as $r$,
implying that the amount of gas in the volume goes as $dm_x \propto r$ as well.
Therefore, the X-ray intrinsic emission goes as
\begin{equation}
L_{xi} =n_e n_ p V \Lambda \propto \left( \frac{dm_x}{V} \right)^2 V \propto r^{-1},
\label{eq:lxem}
\end{equation}
where $\Lambda$ is the emissivity,
and $n_e$ and $n_p$ are the electron and proton number densities, respectively.
The $L_{xi} \propto r^{-1}$ variation is assumed by P09.
There are small variations to this dependance.
A06 considered the relative motion of the two stars, through its influence on the
ram pressures of the two winds. They found that before periastron the X-ray emission
is larger than that given by equation (\ref{eq:lxem}), while after periastron it is lower.

The relation $L_{xi} \propto r^{-1}$ is generic to X-ray emission from colliding fast winds.
For $\eta$ Car the eccentricity is very high, and the increase in the internal X-ray
emission from apastron to periastron is by a factor of $\ga 20$.
Any model, and any observer from what ever direction, would detect a sharp increase in the
X-ray emission when the system is near periastron.
The direction to the observer would change the observed flux because of the
dependance of the column density of the absorbing gas on the direction.
However, there are several unknown parameters of the binary system and the winds
which can be adjusted to accommodate almost any orientation.
For that, the X-ray light curve by itself cannot be used to deduce the periastron longitude,
or even the inclination angle.
This was nicely shown by A06, who could fit the X-ray light curve by using both
an inclination of $i=0^\circ$ (observer above the orbital plane) and $i=42^\circ$
with $\omega=180^\circ$, by slightly varying two parameters of the model
that are related to the two winds (no fine tuning was required).

There is a very deep X-ray minimum lasting for $\sim 10$~week following
periastron passage (Corcoran 2005).
In all models of the kind discussed here it is required to fit the X-ray
emission during the $\sim 10$~week minimum by simply \emph{assuming} this minimum.
Therefore, the behavior of the X-ray emission during the X-ray minimum,
and a few days before and after, cannot be used to discriminate between different
orientations.
P09 reject the $\omega =90^\circ$ semi-major orientation ($\theta=180^\circ$ in their notation)
based only on the behavior during the X-ray minimum.
They presented only the case $\omega = 90^\circ$ with $i = 90^\circ$, rather than taking $i=42^\circ$.

They also examined our preferred case with $(i, \omega)=(42^\circ, 90^\circ)$, and found the
resulted flux at phase $0.98$ to be $\sim 5$ times below the observed one
(Parkin, R., private communication 2009).
However, there are problems with their calculation.
First, P09 assume the emission measure to increase as $EM\propto r^{-1}$, where $r$ is
the orbital separation.
However, the increase in the emission measure is smaller.
For example, at phase $0.99$ the orbital separation is $16$ times smaller than at apastron, but
the emission measure is only $\sim 7$ times larger than the apastron value (H07).
In addition, before periastron the two stars approach each other, an effect that increases the
X-ray emission for constant wind properties, e.g., by $\sim 30 \% $ at phase $0.98$ (A06);
this effect was not considered by P09.

The conclusion is that the secondary wind starts being disturbed weeks before periastron,
and the emission measure does not increase as much as assumed by P09.
To explain the observed flux, the column density cannot increase by the
large factor predicted by the $\omega\simeq 270 ^\circ$ periastron longitude.
A shallower increase in the column density is reproduced by our preferred periastron longitude
$\omega = 90 ^\circ$.

We can summarize this section by stating that fitting the X-ray light curve with a simple model
based on the intrinsic X-ray emission alone cannot teach us
much about the orientation. Almost any inclination and periastron orientation can
be fitted with some adjustment of the poorly known binary and wind parameters.

\section{THE COLUMN DENSITY}
\label{sec:N_H}

In this section we discuss the hydrogen column density as deduced
from X-ray observations by H07.
This section is based on the calculations of KS08,
but significant improvements are added.

We start by describing the observations which we are later rely upon.
The column density toward the hot gas was deduced by H07 using their
XMM-Newton observations.
The binary system itself and its close surroundings are not resolved, but
they are distinguished from the extended X-ray emission in the XMM-Newton
observations.
The XMM-Newton observations cover a very small fraction of the orbit.
The RXTE observations, on the other hand cover more than 2 orbits (Corcoran 2005),
but do not have the spatial resolution to separate the central source from
the extended source.
In any case, most of the X-ray emission during most of the orbit result from
the binary system.



We shall focus our discussion on the hot $>5 \keV$ component.
At phase $0.47$ the column density toward this component is $N_H=1.7\pm 0.3 \times 10^{23} \cm^{-2}$,
as deduced by XMM-Newton observations (H07).
All the $N_H$ observations and their errors as stated by H07 appear in Fig. \ref{fig:NH}.
Though the compact object dominates the emission, the uncertainty can be somewhat
larger due to combination of multiple spectral components in the data.
Our confidence in the column density values derived by H07 is based also on their finding
that the column density has not varied between phases $0.47$ and $0.923$.


The binary parameters are as in our previous paper (where references are given):
The assumed stellar masses are $M_1=120 M_\odot$ and $M_2=30 M_\odot$,
the eccentricity is $e=0.9$, and the orbital period is $P=2024 \days$.
The stellar winds' mass loss rates and terminal velocities are
$\dot M_1=3 \times 10^{-4} M_\odot \yr^{-1}$, $\dot M_2 =10^{-5} M_\odot \yr^{-1}$,
$v_{\rm 1,\infty}=500 \km \s^{-1}$ and $v_{\rm 2,\infty}=3000 \km \s^{-1}$.
For these parameters, the half opening angle of the winds collision region (WCR)
near apastron is $\phi_a \simeq 60 ^\circ$ (A06).
For the inclination angle we take $i = 42^\circ$.

H07 modeled the variable X-ray emission with two components.
A hot component that explains all the emission above $5 \keV$, and an
extra soft component that contributes to the emission below $5 \keV$.
Near apastron the hot component has a temperature of $kT=3.3 \keV$, while the extra
soft component has a temperature of $kT = 1.1 \keV$.
In KS08 we have studied only the hot component because we know this region comes
from the strongly shocked (perpendicular shock) secondary wind before it suffers any
adiabatic cooling. Therefore, we can safely estimate its location to be close to the
stagnation point of the colliding winds (apex), but somewhat closer to the secondary.
The source and location of the $kT \sim 1 \keV$ gas, on the other hand, is
less secure, but it probably resides in an extended region.
The very important thing we do learn from
H07 analysis of the X-ray emission by the $kT \sim 1 \keV$
gas is that the contribution of absorbing gas around the winds interaction region
to the column density is $N_{H-e} \la 5 \times 10^{22} \cm^{-2}$.
This implies that any column density of $N_H >  5 \times 10^{22} \cm^{-2}$ must
come from material close to the binary system, at most $\sim 100 \AU$.

P09 tried to fit the RXTE light curve of the emission in the $2-10 \keV$ band (Corcoran 2005).
The temperature of the extra soft component is $kT_{\rm soft} \simeq 1.1 \keV$,
and its contribution to the $2-10 \keV$ band is expected to be small relative
to that of the hot ($kT=3.3 \keV$) component (e.g., Sarazin \& Bahcall 1977).
Indeed, the hot component contributes $70 \%$ of the observed X-ray emission
in the $2-10 \keV$ band (deduced from figures 8 and 9 of H07).
Therefore, in fitting the column density based on the X-ray emission in the
$2-10 \keV$ band, one better compare it to the column density toward to
hot component of H07 (their table 5), rather to the entire spectrum of H07
(the entire spectrum studied by H07 goes below $2 \keV$).
In any case, we shall stay with the hot component as it is better defined.

The column density toward the hot gas at phase $0.47$ is
$N_H=1.7\times 10^{23} \cm^{-2}$. This is hard to explain in a model where
the secondary is toward us during this phase, i.e., $\omega \sim 270^\circ$.
This is because the opening angle of the conical shell (the collision surface of the
two winds) is $\phi_a \simeq 60^\circ$, while the inclination angle is $i\simeq 42^\circ$.
This implies that $90^{\circ} -i< \phi_a$, such that our line of sight
toward the shocked secondary wind would go through the undisturbed secondary wind.
The column density through this low density gas is very low.
Instead, we suggest that the secondary star resides on the far side during apastron,
and the column density includes the undisturbed
primary wind as well as the shocked primary wind.

\begin{figure}[!t]
\resizebox{0.89\textwidth}{!}{\includegraphics{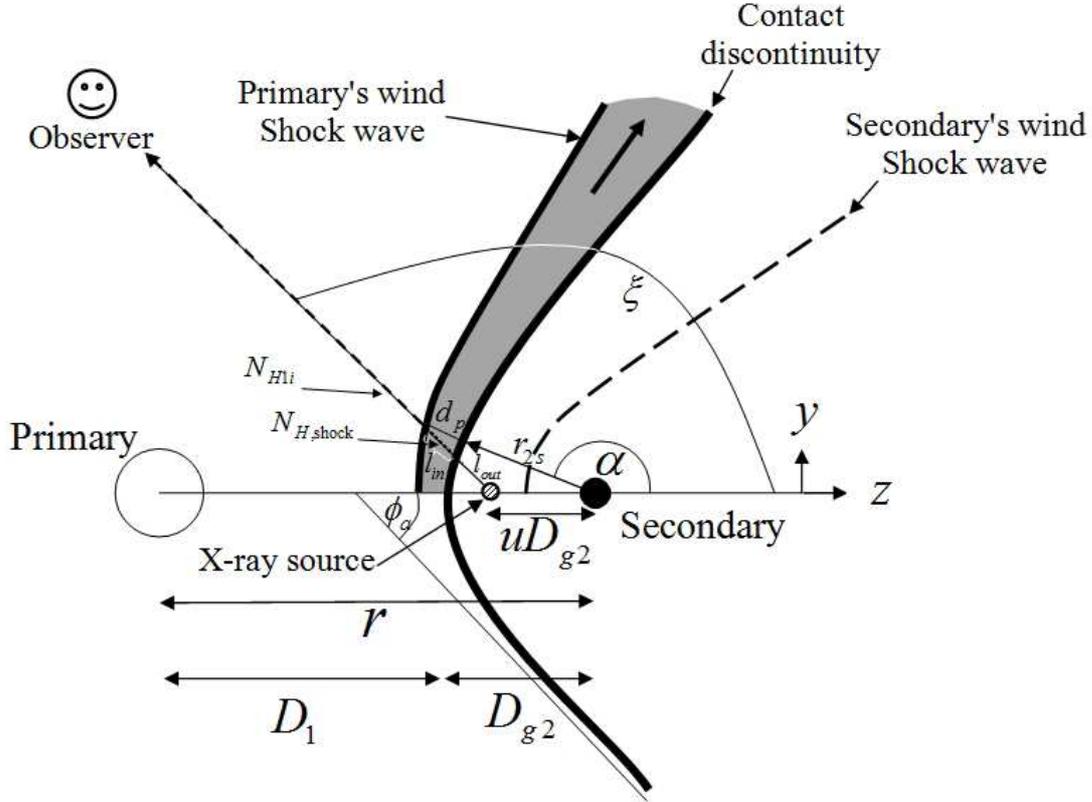}}
\caption{\footnotesize A schematic cut through a plane perpendicular to the
orbital plane, showing the winds collision region (WCR) of the two stellar
winds at apastron, where the WCR is not rotated and its symmetry axis consolidates with
the line connecting both stars (i.e. $\delta \phi = 0$).
Several quantities used in the paper are defined.
The point marked `X-ray source' in the post shocked secondary
wind where the hot gas that is the source of the emission above $5 \keV$ resides.
{}From there we calculate the value of $N_H$.
Approaching periastron the geometry becomes more complicated (see Fig
\ref{fig:hyperboloid}). }
\label{fig:NHgeo}

\end{figure}

To calculate the column density for our preferred orientation of  $\omega=90^\circ$,
we follow KS08 and use the geometry as drawn schematically in
Fig. \ref{fig:NHgeo}.
The contact discontinuity shape is approximated by an hyperboloid at
a distance $D_{g2}\simeq0.3 r$ from the secondary at the stagnation point
(the stagnation point is the point where the two winds' momenta balance each other).

Our model of the primary wind includes a weak pre-shock magnetic field. In the
post shock region the gas is highly compressed, and the magnetic field becomes dominant.
The magnetic pressure limits the compression of the postshocked primary wind.
The uncertainties in the intensity and geometry of the magnetic field
introduce large uncertainties (Kashi \& Soker 2007a).
The magnetic field role is parameterized by the compression factor $\eta_B$
which is the ratio between the pre-shock magnetic and ram pressure
(see equations 9-11 in Kashi \& Soker 2007a).
Following the model for the post shocked primary wind which had been presented
in Kashi \& Soker 2007a, we will use the quantity
\begin{equation}
\label{eq:fm}
f_m =
 \begin{cases}
v_{\rm wind1} \sin^2\psi                       & \qquad  v_{\rm wind1} \sin^2\psi < \left(\frac{3}{\eta_B \sin^2\psi}\right)^{1/2}
    \\
\left(\frac{3}{\eta_B \sin^2\psi}\right)^{1/2} & \qquad  v_{\rm wind1} \sin^2\psi < \left(\frac{3}{\eta_B \sin^2\psi}\right)^{1/2}
 \end{cases}
\end{equation}
where $\psi$ is the angle between the slow wind velocity and the
primary wind shockwave, $v_{\rm wind1}$ is the relative speed between the secondary
and the primary wind, given in equation (\ref{eq:vwind1}) below,
and $\eta_B=0.002$ is a parameter (Kashi \& Soker 2007a).

The value of $\eta_B=0.002$ implies that the magnetic field has a negligible
role before the shock, but not after.
The column density is
\begin{equation}
n_H = \frac{0.43f_m \dot M_1}{4 \pi r_{1s}^2 v_{\rm wind1} \mu m_H } .
\label{eq:nH}
\end{equation}
The compression factor together with a few more
assumptions allow us to calculate the velocity of the post shock primary wind
out from the shock region, and the width of the conical shock, $d_p$
(see equation 12 in Kashi \& Soker 2007a).
{}From the width we can calculate the values of $l_{out}$ and $l_{in}$,
defined in Fig. \ref{fig:NHgeo}.

The source of the hard X-ray is the post-shocked secondary wind
(Pittard \& Corcoran 2002; Corcoran 2005; A06), taken here at the point
marked on Fig. \ref{fig:NHgeo} by `X-ray source'.
This point is located at a distance of $(1-u)D_{g2}$ from the
stagnation point, or a distance of $uD_{g2}$ from that point.
The X-ray emitting region is more extended, but it does not extend to large
distances from the secondary, and our assumption is adequate.
We assume $u = 0.7$.
As evident from the figure, the column density has
two main components: The post-shocked primary wind component,
$N_{H,\rm{shock}}$ (the conical shell), and the undisturbed,
free-expanding, primary wind component ($N_{Hi1}$). We calculate the
contribution of each component to the total column density
($N_{H,\rm{tot}}$) as a function of orbital angle $\theta$
($\theta$=0 at periastron).

As in our previous papers, we approximate the shape of the colliding
winds conical shell as an hyperploid.
In the present study we consider some additional effects, listed next,
that makes the geometry more complicated and the results more accurate.
Similar considerations were used by us (Kashi \& Soker 2009b) to explain
the complex P Cygni profile of the He~I~$\lambda 10830${\AA} high
excitation line.

(1) \textbf{Wind acceleration.}
We take the primary wind acceleration into account.
We describe it as a $\beta$-profile
\begin{equation}
v_1(r_1)=v_s+(v_{\rm 1,\infty}-v_s)\left(1-\frac{R_1}{r_1}\right)^{\beta} ,
\label{eq:v1}
\end{equation}
with a parameter $\beta=1$, where $v_s=20 \km \s^{-1}$
is the sound velocity on the primary surface,
$v_{\rm 1,\infty}=500 \km \s^{-1}$ is the primary wind
terminal velocity (which was already defined earlier),
$r_1$ is the distance from the primary center,
and $R_1 = 180 \AU$ is the primary radius.
The acceleration of the primary wind becomes important close to periastron,
when the binary separation becomes only slightly larger than the primary radius.
Being slower and denser close to the surface, the accelerating primary wind
yields a denser post shocked primary wind,
and affects other variables in this already complex geometry.
Its slower velocity makes the hyperboloid asymptotic opening angle wider,
and its rotation more pronounced, as we now explain.

(2) \textbf{Hyperboloid asymptotic opening angel.}
The radial (along the line joining the two stars) component of the
relative velocity between the secondary star and the primary wind is
$v_1-v_r$, where $v_r$ the radial component of the orbital velocity;
$v_r$ is negative when the two stars approach each other.
The relative speed between the secondary and the primary wind is
\begin{equation}
v_{\rm wind1} = \left[v_\theta^2 + (v_1-v_r)^2 \right]^{1/2},
\label{eq:vwind1}
\end{equation}
where $v_\theta$ is the tangential component of the orbital velocity.
The orbital motion and the variation of the primary wind velocity with distance
from the primary have an influence on the hyperboloid asymptotic opening angel $\phi_a$.
We use the expression given by Eichler \& Usov (1993)
\begin{equation}
\phi_a \sim 2.1
\left(1-\frac{\eta_w^{{4}/{5}}}{4}\right)\eta_w^{{2}/{3}} ,
\label{eq:phia}
\end{equation}
where
\begin{equation}
\eta_w \equiv \sqrt{\frac{\dot M_2 v_{\rm 2,\infty}}{\dot M_1 v_{\rm wind1}}}
\label{eq:eta}
\end{equation}
(note that this definition of $\eta_w$ is different from the one used by P09, where
$\eta_{P09}=\eta_w^2$).
$\phi_a$ has a minimum value of $\sim 58^\circ$ $33$~days before periastron,
and reaches maximum of $\sim 72^\circ$ $25$~days after periastron.

\begin{figure}[!t]
\resizebox{0.89\textwidth}{!}{\includegraphics{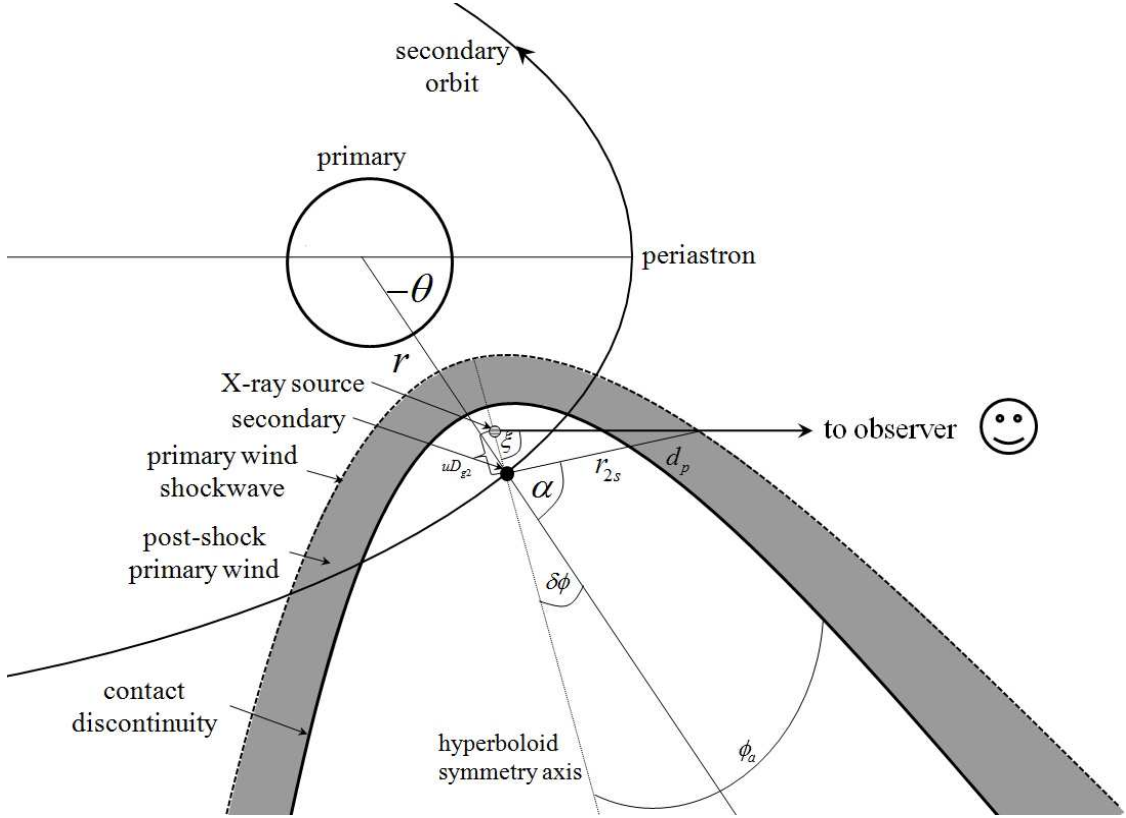}}
\caption{\footnotesize The geometry of the binary system and the conical shell
in the orbital plane.
Several quantities are defined on the figure.
We approximate the shape of the colliding winds as an hyperploid, with an
asymptotic opening angel $\phi_a$.
$\delta \phi$ is the angle between the symmetry axis of the hyperboloid
near the stagnation point, and the line joining the two stars. }
\label{fig:hyperboloid}
\end{figure}

\begin{figure}[!t]
\resizebox{0.89\textwidth}{!}{\includegraphics{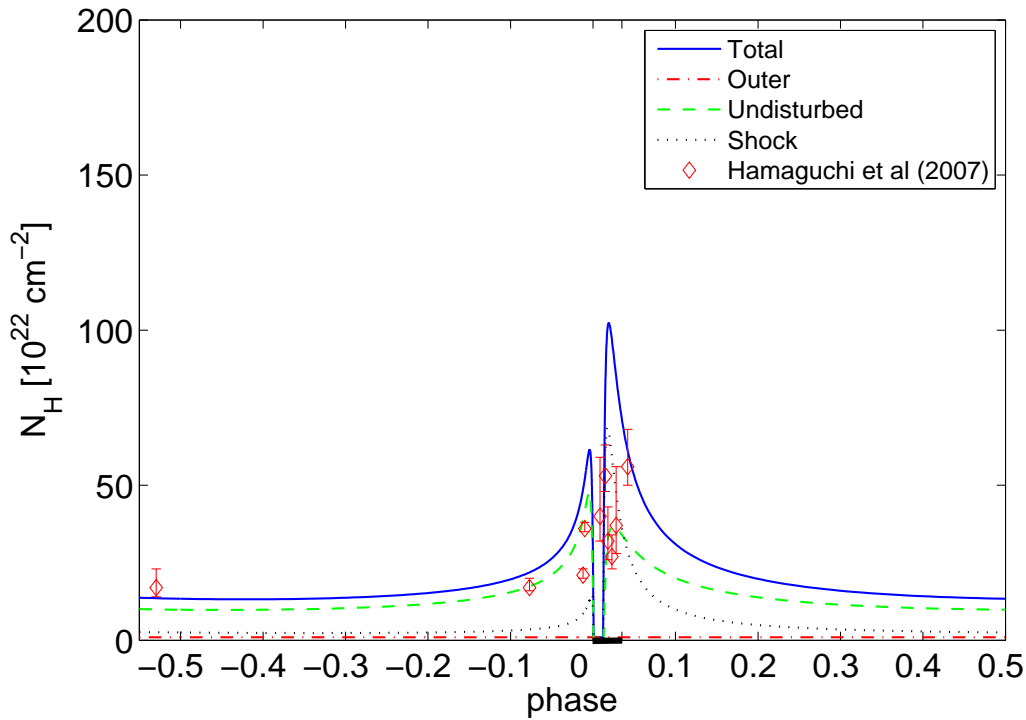}}
\caption{\footnotesize The column density toward the hot gas
as obtained from our model, $(i, \omega)=(42^\circ, 90^\circ)$.
The dashed-green line represents the undisturbed
(free-expanding) primary wind component ($N_{Hi1}$); The dotted-black
line represents the post-shocked primary wind component ($N_{H,\rm{shock}}$);
The dot-dashed red line represents the constant column density from the Homunculus
and ISM.
The solid-blue line is the sum of all three components.
The $N_H$ toward the hot gas from Table 5 of H07 is plotted as red diamonds.
The thick line on the horizontal axis mark the accretion period, during which the
calculation of $N_H$ is not applicable.}
\label{fig:NH}
\end{figure}

(3) \textbf{Hyperboloid rotation.}
We take into consideration the rotation of the hyperboloid relatively to the line
connecting the two stars.
Namely, the winding of the WCR around the secondary star,
as the secondary orbits the primary across periastron.
This rotation (winding) has a considerable influence close to periastron.
We define $\delta \phi$ to be the angle measured from the secondary between the
direction to the primary and that to the stagnation point (see Soker 2005 for further details)
\begin{equation}
\cos (\delta \phi)=\frac{v_1-v_r}{v_{\rm wind1}}.
\label{eq:deltaphi}
\end{equation}
We find that the maximum value of $\delta \phi$ is obtained $\sim 6$~days after periastron,
where it reaches $\simeq 64^\circ$.
This makes the rotation of the hyperboloid non-negligible and very important
for the more accurate calculation we perform in this paper.
The equatorial plane of the geometry described above is presented in Fig. \ref{fig:hyperboloid}.

We take the inclination angle to be $i=42^\circ$, and assume that
the secondary is away from us during a apastron passage ($\omega=90^\circ$).
This geometry explicitly determines the direction from which the system is observed
(i.e. line of sight) at each orbital phase.
For every orbital angle $\theta$ we calculate the relevant
direction angle $\xi$ to the observer. Considering the orientation of
the conical shell at that orbital angle, we calculate the thickness of
the conical shell in that direction and integrate $n_H$
over the width to find the column density of the first component:
\begin{equation}
N_{H,\rm{shock}} = \int_{l_{\rm{out}}}^l n_H\,dl \label{eq:NHshock},
\end{equation}
where $l=l_{\rm{out}}+l_{\rm{in}}$ (see Fig. \ref{fig:NHgeo}).
The second component contributing to the column density, the undisturbed primary wind,
is calculated from the point on the line of sight where the shock terminates,
to infinity (contribution decreases fast with distance)
\begin{equation}
N_{Hi1} = \int_l^\infty n_H\,dl \label{eq:NHi1}.
\end{equation}
To calculate the total column density we added a constant third component of
$1 \times 10^{22} \cm^{-2}$ to account for the material residing in the outer regions,
e.g., in the Homunculus and in the ISM, the same value as P09 used.

The three column density components and the total column density are
plotted in Fig. \ref{fig:NH}. The column density toward the hot component
based on the spectrum above $5 \keV$, $N_H[>5 \keV]$,  from H07 is also plotted.

Our model reproduces to within a factor of two the results of
H07, and is with agreement with its qualitative behavior.
This is done without any parameters fitting.
Namely, we simply take the values of the different parameters as in our previous papers,
where some parameters were adjusted to fit other observations of $\eta$ Car, such
as the radio emission, some helium lines, and more.

Our results clearly shows that from our preferred line of sight
($\omega=90^\circ$,$i=42^\circ$) the column density hardly changes during
most of the orbital cycle and can supply the required high column density,
in accord with observations. When the system approaches periastron passage
there is a fast increase of $N_{H,\rm{shock}}$ and $N_{Hi1}$, followed by a
decrease after periastron passage.

\begin{figure}[!t]
\resizebox{0.89\textwidth}{!}{\includegraphics{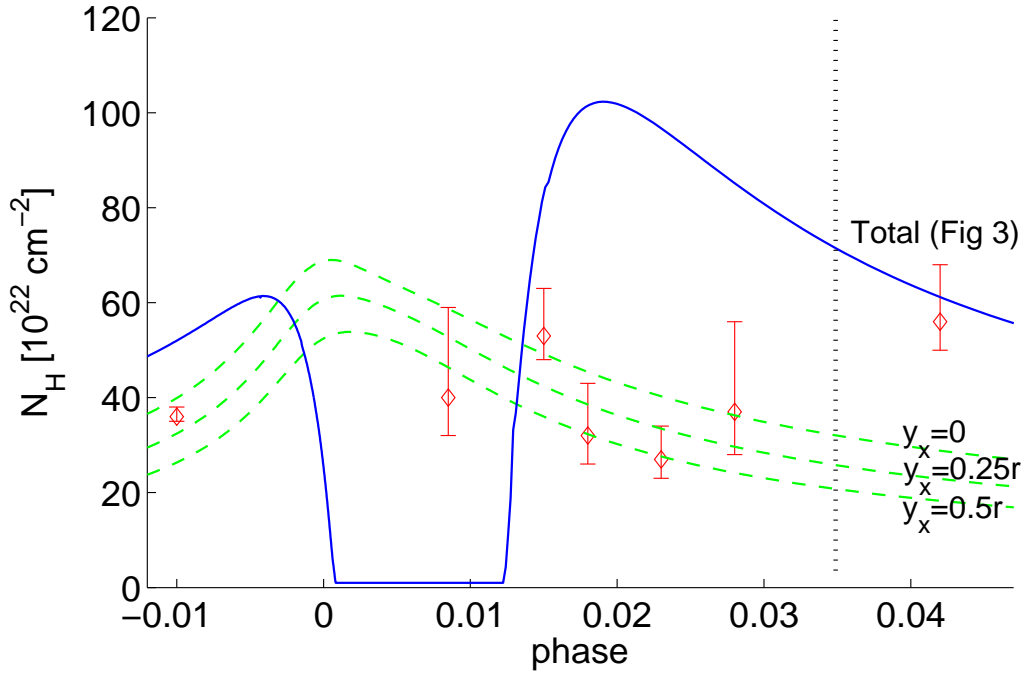}}
\caption{\footnotesize The column density toward the hot gas
as obtained from our model, $(i, \omega)=(42^\circ, 90^\circ)$ when
no winds collision occurs (3 dashed lines).
The column density is due to the spherically symmetric undisturbed primary wind.
The vertical line at phase $0.035$ marks the end of the X-ray minimum.
The observation point at phase $0.042$ is better fitted with
a model that includes the conical shell (Fig. \ref{fig:NH}) than with a spherical primary wind,
and is shown as the solid blue line, relevant only after phase $0.035$.}
\label{fig:min}
\end{figure}

One thing must be kept in mind.
The ten-week X-ray minimum cannot be explained by the model.
A different ingredient must be incorporated; one that extinguishes the conical shell.
We take this process to be accretion onto the secondary star.
For that, our calculation of $N_H$ with the WCR included is not applicable at
all during the minimum, i.e., the phase period $\sim 0-0.035$.
The flow structure around the secondary is more complicated, as it includes the accretion
process, and possibly a polar outflow (jets) that results the extra soft source during the
X-ray minimum (see section \ref{sec:ex}).
The column density during the minimum might be better represented by an undisturbed
primary wind in the entire space.
As we are not sure where the polar outflow is shocked and forms the X-ray emitting regions,
we take 3 possibilities for its location: At the location of the
secondary itself (which is the average of two opposite jets);
a region above the primary (perpendicular to the equatorial plane)
of $y_x=0.25r$, and of $y_x=0.5r$.
In Fig. \ref{fig:min} the column density to the X-ray emitting region
in the cases is plotted during the X-ray minimum, and $20$ days (a phase of $0.01$)
before and after the minimum.

We note that during the X-ray minimum the column densities toward the hard and soft
X-ray sources are about equal (ratio of $\sim 1$), compared to a ratio of $> 3$ near
apastron. The results presented in Fig. \ref{fig:min} suggest that the hard and soft
X-ray components reside close to the secondary, and are not originated in the tail of the WCR.
We also note that the observation point at phase $0.042$ is better fitted with
a model that includes the conical shell (Fig. \ref{fig:NH}) than with a spherical primary wind,
also shown as the solid blue line in Fig. \ref{fig:min}.
This points is well after the X-ray minimum, and this is an expected result.
The point at phase $-0.01$ can be marginally fitted if the WCR does exist,
by changing somewhat the unknown parameters.
However, it seems we can better fit it by assuming that the WCR is already
highly disrupted at that time.
Namely, the collapse of the WCR starts around $-0.01$ or somewhat earlier, as suggested by A06.

We note that the observed column density at phase $0.47$ of
$N_H=17\times10^{22} \cm^{-2}$ (H07) occurred when the $2-10 \keV$ emission
was $10-15 \%$ above its average value during that time (Corcoran 2005).
This could result from a higher density of the primary wind.
According to the model of A06 the dependance of the X-ray emission on the primary
mass loss rate is $L_x \sim \dot M_1^{1/2}$. Namely, it is quite possible that the
primary mass loss rate, and hence $N_H$, were higher than the average value near apastron
by a factor of $\sim 1.25$, and this is the reason H07
did not find the column density to increase between phases
$0.47$ ($-0.53$) and 0.92 ($-0.08$).

Another effect not considered by us, and that introduces more variations,
both in the time variation and in the absorption by the conical shell
along different directions, is the corrugated structure of the shocked primary
wind that results from instabilities (Pittard \& Corcoran 2002; Pittard et al. 1998;
Okazaki et al. 2008; P09).

As mentioned earlier, from behind the secondary shock (namely, if
the secondary is toward us near apastron), it is not possible
to account for $N_H=17\times10^{22} \cm^{-2}$ column density near apastron.
It cannot come from the nebula, as the nebula can supply
$5 \times10^{22}\cm^{-2}$ at most, as deduced from the column density toward
the low temperature gas (the extra soft component, H07).
Also, from figure 2 of P05 we learn that when the system is at apastron,
there is a region extending to $\sim 500 \AU$ behind he secondary star that
is cleaned from the primary wind by the tenuous secondary wind.
Namely, when observed through the secondary wind region the column density of the
primary wind is very low, practically negligible.
In the case of $\omega\simeq 270 ^\circ$ (which we oppose), one would indeed observe
through the secondary wind, because $90^\circ -i < \phi_a$.

\begin{figure}[!t]
\resizebox{0.89\textwidth}{!}{\includegraphics{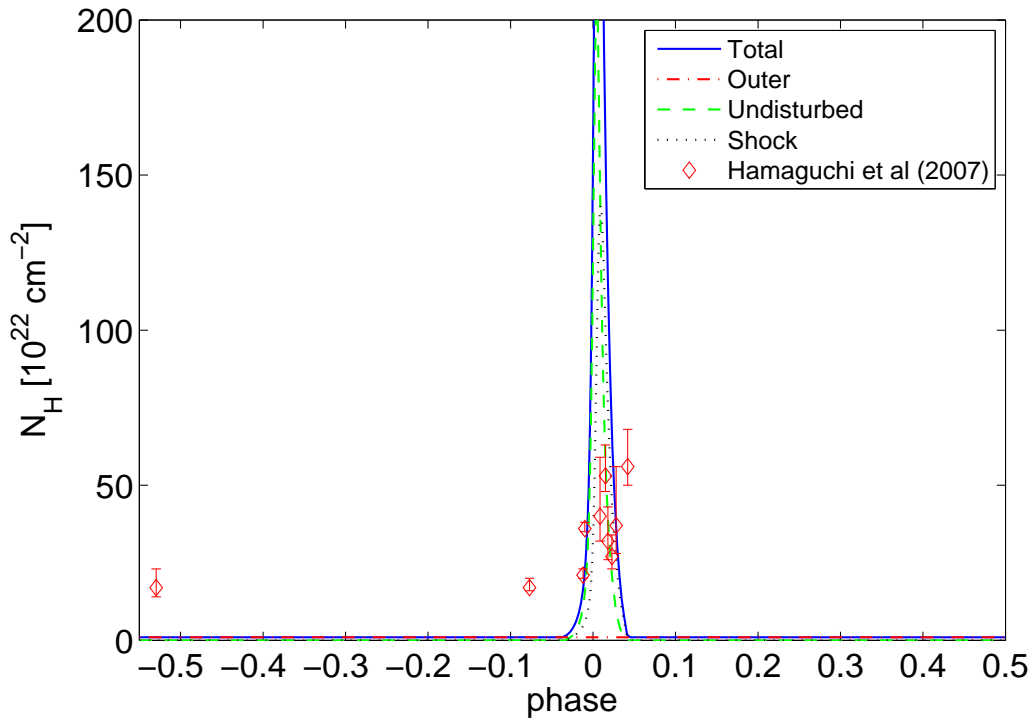}}
\caption{\footnotesize The column density obtained using $\omega=270^\circ$,
namely, the secondary is toward us near apastron.
Most of the binary orbit an observer at $\omega=270^\circ$ would observe
the system through the secondary fast wind, which has negligible contribution to
the column density.
Using this viewing angle from behind the shock at apastron, it is not possible
to account for the observed column density $N_H=17\times10^{22} \cm^{-2}$.}
\label{fig:NH270}
\end{figure}

To show this point, we have repeated our calculation for an
opposite periastron longitude $\omega=270^\circ$.
The results are shown in Fig. \ref{fig:NH270}.
We see that for that case the column density long before and after periastron
is highly underestimated.
Moreover, close to periastron the calculated hydrogen column density reaches
$\sim 3\times 10^{24}\cm^{-2}$ (outside the plotted region of Fig. \ref{fig:NH270}).
This is about 5 times the observed value, while our orientation yields values
less than twice the observed value.
Even though we admit our calculation of $N_H$ is not applicable during the
X-ray minimum, we crudely do get the increase factor of the column density between
apastron and periastron as observed by H07.
In the case of $\omega=270^\circ$, our calculations show a huge jump in the column density,
not compatible with observations.
This is mostly pronounced at phase $0.923$ ($-0.077$), when the calculated column density
for $\omega=270^\circ$ is too low to explain the observed value, and the minimum has not started yet.

\section{THE COLLAPSE OF THE WIND COLLISION REGION ONTO THE SECONDARY STAR}
\label{sec:collapse}

P09 discussed the possibility that the WCR collapses onto the secondary.
In their discussion they mentioned that if the collapse occurs, then
the secondary wind is continued to be blown on the other side of the secondary.
They do not consider accretion.
We now show that the regular secondary wind cannot be blown during the event, not
even half of it, and that if the WCR collapse occurs, then accretion in inevitable.

\subsection{The residual wind}
\label{sec:residual}
P09 assumed that during X-ray minimum about half of the secondary wind is continued
to be blown at $\sim 3000 \km \s^{-1}$ from the side of the secondary not facing
the primary.
They then took the X-ray to come from regions away from the secondary, as if
the colliding wind regions was unaffected. We show that this cannot be the case.

Let the wind blown posses a fraction $\zeta$ of the regular wind, such that the mass
loss rate is $\dot M_{2m} = 10^{-5} \zeta M_\odot \yr^{-1}$.
{}From figure 3 of P09 we see that during the X-ray minimum the conical shell is
wind-up in such a way that the secondary wind cannot escape.
At phase 1.02 (40 days after periastron passage) the wind can propagate at most $20 \AU$
before encountering the primary dense wind; most of the secondary wind is shocked at
a much closer distance of $< 5 \AU$.
It does it within $20 \AU/ 3000 \km \s^{-1} = 12$~day, which is shorter than the
time since minimum starts, but not by much.
The volume of space available for the shocked secondary wind is
$\Delta V  \sim (20\AU)^3 \simeq 3 \times 10^{43} \cm^{3}$.
The mass blown during this time (40 days) is $\Delta M=1.1 \times 10^{-6} \zeta M_\odot$.
The proton number density is
\begin{equation}
n_p \simeq  3 \times 10^7 \zeta \cm^{-3}.
\label{eq:np}
\end{equation}
At a temperature of $10^8 \K$ the emissivity is $\Lambda =2.5 \times 10^{-23} \erg \s^{-1} \cm^3$,
and the total X-ray luminosity of the bubble is (about half emitted in the $2-10 \keV$ band)
\begin{equation}
L_{xm} \simeq 2 \times 10^{35}
\left( \frac{\zeta}{0.5} \right)^2  \erg \s^{-1}.
\label{eq:lx}
\end{equation}
The emission measure ${\rm EM} \equiv n_p n_e \Delta V$, where $n_e$ is the electron density, is
\begin{equation}
{\rm EM}_{xm} \simeq  10^{58}\left( \frac{\zeta}{0.5} \right)^2  \cm^{-3}.
\label{eq:em}
\end{equation}
This emission measure is an order of magnitude larger than that of the hot gas
deduce from X-ray observations during minimum X-ray emission (H07).
If we reduce $\zeta$ much below $\zeta=0.5$, the volume of the shocked secondary wind will
be smaller. So in order to reduce the emission measure to
${\rm EM}_{xm} \simeq 3 \times 10^{57} \cm^{-3}$, we should have $\zeta < 0.1$.
Namely, an almost complete shutdown of the secondary wind.

As seen in figure 2 of P09, the volume available for the shocked secondary wind
during X-ray minimum is not closed, and the post-shocked wind will flow out, reducing the
density and emission measure.
However, the outflow time scale is not much shorter than the duration of the minimum,
and this will not reduce the emission measure by much.
On the other hand, there are effects that operate to increase the emission measure in
the model of P09.
(1) The side that blows the wind in their model is opposite to the primary direction.
However, because of the winding of the tail (its spiral structure) this side is not toward
the tail of the shocked region, and the secondary wind
will be shocked in a short distance from the secondary of only several$\times \AU$ on average,
at phase 1.02.
(2) As mentioned above, the volume of $\Delta V \simeq (20\AU)^3$ is taken from their
figure 2, which is drawn for a full blown wind. If the secondary wind is weaker, the volume
will be smaller.
(3) As discussed below, the gravity of the secondary will bend the
primary wind stream lines toward the secondary. This will increase the ram
pressure of the primary wind, and by that further reduce the volume.

\subsection{The accretion phase}
\label{sec:acc0}

\subsubsection{The inevitability of accretion}
\label{sec:acc}

P09 mentioned that the primary wind comes very close to the secondary, but did not
consider accretion according the Bondi-Hoyle-Lyttleton model, but rather took the accretion
radius to be the secondary radius. This is not consistent with the well established
accretion process where the accretion radius should be the Bondi-Hoyle accretion radius
\begin{equation}
R_{\rm acc2} = \frac {2G M_2}{v^2_{\rm wind1}}= 0.2
\left( \frac{v_{\rm wind1}}{500 \km \s^{-1}} \right)^{-2}  \AU,
\label{eq:racc2}
\end{equation}
for a secondary mass of $M_2=30 M_\odot$.
The accretion radius was calculated by A06, and was found to be
$R_{\rm acc2} \simeq 0.5 \AU$ at phase $1.01-1.02$.
This is $\sim 5$ times the radius of the secondary star.

As was discussed by Soker (2005), A06, and Kashi \& Soker (2009a), the
secondary stellar radiation pressure and wind cannot prevent accretion.
We note that Soker (2005) considered the ram pressure of the secondary wind,
as well as the radiation pressure.
That ram pressure and radiation pressure cannot prevent accretion is true for
a smooth wind, and more so as dense blobs are expected to exist
in the primary wind due to stochastic mass loss processes and instabilities in the
colliding winds (Pittard \& Corcoran 2002).
P09 mentioned that the secondary wind can destroy the falling dense blobs.
This process should be studied in a future paper, but must include the magnetic
field within the blobs. The magnetic field in the blob might prevent efficient ablation.
Kashi \& Soker (2009a) further include the acceleration zone of the primary wind,
and find the secondary stellar gravity to be more important even than
what was found by Soker (2005) and A06.
Neglecting the secondary gravity by P09 makes their results
questionable, e.g., their claim that no substantial accretion occurs is not supported.

We do note that the accreted mass cannot account for the
strong X-ray component by itself.
The temperature of the strong soft X-ray component during the X-ray minimum
is $kT \simeq 0.5-1 \keV$ (H07), about an order of magnitude below the hard
X-ray component.
This temperature is formed in the postshocked region of gas flowing with a velocity of
$650-1300 \km \s^{-1}$.
In the case of an inflow the gas is compresses, and a velocity range of
$\sim 600-1200 \km \s^{-1}$ is required.
In the case of an outflow, adiabatic cooling occurs and the
required outflow velocity is $\sim 800-1600 \km \s^{-1}$.
The free fall velocity onto the secondary star is
$v_{\rm ff} = 750 (M_2/30 M_\odot)^{1/2}(R_2/20 R_\odot)^{-1/2} \km \s^{-1}$.
However, the accreted flow will be shocked somewhere above the surface, and it seems it
cannot account for the temperature of the soft component.

Neither the accreted mass can account for the emission measure.
The high emission measure of the soft component is
$EM_s=n_e n_p V \simeq 10^{59} - 10^{60} \cm^{3}$ (H07),
for an average of $\sim 10^{59.5} \cm^{3}$.
It is not clear how long this high emission measure phase lasts, as H07 give only two
measurements during the X-ray minimum. We assume this high emission measure phase to last 5 weeks.
During a time period of $t_m \simeq 5$~weeks the total intrinsic X-ray emission (before absorption)
is  $E_{xs}=EM_s \Lambda t_m \simeq  5 \times 10^{43} \erg$.
In Kashi \& Soker (2009a) we estimated the accreted mass during the $10~{\rm weeks}~=0.2 \yr$
X-ray minimum to be $M_{\rm acc} \simeq  0.4 - 3.3 \times 10^{-6} M_\odot$.
The available energy is at most $E_a \simeq G M_2 M_{\rm acc}/R_2 \simeq 10^{43} \erg$,
for $M_{\rm acc} = 2 \times 10^{-6} M_\odot$.
The accreted gas cannot account for the high emission measure of the soft X-ray component.
An alternative is discussed in section \ref{sec:ex}.

\subsubsection{X-ray emission during the accretion phase}
\label{sec:ex}

In our calculations (here and in KS08) we studied the properties of the hot
gas, i.e., the component that is the source of the X-ray emission above $5 \keV$
according to H07.
Let us examine the properties of the extra soft component.
Its temperature is $kT \simeq 0.5-1.1 \keV$, and its contribution to the
X-ray emission is larger than that of the hot component in the $<3.2 \keV$ X-ray band (H07).
In our calculations (here and in KS08) the column density is calculated
to a region near the stagnation point of the colliding winds.
This is the region where the secondary wind shock is strong and no adiabatic
cooling has occurred yet, and hence hard X-ray emission is expected.
In all cases and parameters the average calculated column density during the
X-ray minimum is above the observed value.
This is true for both the hot and the extra soft component.
This suggests that the X-ray emission region is located at somewhat larger
distance from the secondary star.

During the minimum the emission measure of the hard X-ray decreases
substantially, while that of the soft X-ray increases by more than an order of magnitude (H07).
As the emission measure of the soft component increases, so does the column density
toward its emission region. The column densities of the hard and soft component become
comparable; at all other phases the column density of the soft component is smaller than
that of the hard component.
The high column density shows that the emitting region cannot be too far
from the secondary star.
We suggest that the regular winds collision does not occur during the X-ray minimum.
Rather, there is a polar outflow (jets).

During the X-ray minimum the regular secondary wind is substantially
suppressed to only few percents of its regular intensity.
This explains the small emission measure of the hard component.
Instead, we suggest that the accretion disk around the secondary star,
formed by the accreted primary wind material, forces the outflow to
direct in the polar directions. The outflowing material is composed of the
secondary wind and accreted gas.
In Kashi \& Soker (2009a) we estimated the average rate at which the
secondary accretes mass from the primary during the X-ray minimum to be
$\dot M_{\rm acc} \sim 10^{-5} M_\odot \yr ^{-1}$.
We note that the accretion rate is about equal to the undisturbed secondary wind.
This suggests that the accretion process substantially disturbes the wind,
but cannot completely prevent an outflow.

To account for the soft X-ray component, with a total radiated energy of
$E_{xs} \simeq  5 \times 10^{43} \erg$ (see section \ref{sec:acc}), we
can take an average mass loss rate over the 10~weeks of
$\dot M_{\rm polar} \sim 10^{-5} M_\odot \yr^{-1}$ with a velocity of $1600
\km \s^{-1}$.
P09 found that the post shocked region of gas flowing at this velocity
can match the spectrum at X-ray minimum.
Namely, the secondary mass loss rate does not change much, but it is forced to the
polar direction. The high mass loss rate over a smaller angle results in a
less efficient acceleration process, and the terminal speed is about a half of
its regular value.
This polar outflow cannot escape from the dense primary wind, and most of
its kinetic energy is converted to X-ray emission.
The presence of a polar outflow in $\eta$ Car, before and during the X-ray
minimum, was
suggested before, based on the behavior of the He~II~$\lambda$4686 line
(Soker \& Behar 2006),
and on the Doppler shift of X-ray lines (Behar et al. 2007).
Theoretical motivation for a strong secondary polar outflow during and at
the end of the X-ray minimum
is given by Kashi \& Soker (2009a).

\section{DISCUSSION AND SUMMARY}
\label{sec:diss}

We studied the usage of the X-ray properties to deduce the orientation
of the semi-major axis of $\eta$ Car.
It is agreed by most researchers that the inclination of the binary system
is $i \simeq 42^ \circ$. However, there is a dispute on the direction of the
semi-major axis in the equatorial plane, the so called periastron longitude,
or orientation.
The definition of the periastron longitude angle in the orbital plane is
given in equation (\ref{eq:omega}).

In section \ref{sec:xray} we argued that the X-ray light curve by itself
cannot be used to deduce the orientation. This conclusion is based in part on the results
of A06, who showed that with some adjustment of the binary unknown parameters, e.g.,
exact wind properties, the X-ray light curve can be fitted by using almost
any periastron longitude and any inclination angle.
The only large differences between the different orientations occur near periastron passage.
But there, all models must assume a suppression of X-ray emission, such that
the period of $\sim 3$ months around periastron passage is useless in simple models
that use only the X-ray light curve.

In section \ref{sec:N_H} we modelled the column density as deduced
from X-ray observations by H07.
We considered some additional effects that we didn't take into account in
KS08, which make our modelling more accurate and our conclusion that
$\omega=90^\circ$ more reliable.
The periastron longitude deduced in P09,
$\omega=270^\circ - 300^\circ$, is more or less opposite to our preferred
value of $\omega=90^\circ$.
We showed that their periastron longitude is not in agreement with
observations of the column density, particulary close to apastron.

In section \ref{sec:residual} we considered the secondary wind during the
X-ray minimum.
We showed that the emission measure of the hard X-ray emission as measured
by H07 during the X-ray minimum, constrains the regular secondary wind, i.e.,
that with the same terminal velocity of $\sim 3000 \km \s^{-1}$,
to have a mass loss rate of $< 0.1$ times its regular value, and probably
only $\sim 0.01$ times its regular value.

The emission measure of the soft X-ray component, on the other hand, is very
large during the X-ray minimum (H07).
In section \ref{sec:ex} we suggested that the source of this emission is
shocked polar outflow (or a collimated polar wind, or two jets).
The polar outflow is formed by the accretion process, which focuses the
secondary wind to polar directions, and launches some material from the
accretion disk.
The inevitability of the accretion process (Kashi \& Soker 2009a) was
discussed in section \ref{sec:acc}.
The denser wind along the polar directions makes acceleration less efficient.
This result in a slower outflow $\sim 1000-2000 \km \s^{-1}$, and softer
X-ray emission.

It is important to note that the presence of a polar outflow before and
during the X-ray minimum was suggested before (Soker \& Behar 2006; Behar et al.
2007; Kashi \& Soker 2009a).

To summarize the discrepancy between the model of P09 (who claimed for $\omega\simeq
270^\circ$) and ours ($\omega\simeq 90^\circ$), we think that P09 deduced a
wrong semimajor axis orientation for the following reasons.
\begin{enumerate}
\item They gave a heavy weight to the light curve. However, the light curve
can be fitted by almost any inclination and orientation angles, with some
adjustment of parameters (no fine tuning is required; A06).
\item In the $2-10 \keV$ band the main source ($70 \%$ of the observed flux)
of the X-ray emission is the hot component that was defined by H07.
The column density should be the one toward this component, and not toward
the entire emitting gas, that contains
also component at a temperature of $\sim 1 \keV$.
The column density calculated by P09 is much lower than observed.
\item P09 assumed that the emission measure goes as $r^{-1}$, where $r$ is the
orbital separation.
However, the emission measure deduced from observations increases by a smaller factor
as the system approaches periastron (H07).
\item According to the semimajor orientation deduced by P09, near apastron
our line of sight goes through the tenuous secondary wind.
The column density is very low as evident from our Fig. \ref{fig:NH270}.
The calculation of P09 did not show this low column density.
\item The calculated behavior during the ten-week X-ray minimum
(and few days before and after) cannot be reproduced by the models
(neither ours, or A06, or of P09),
and cannot be used to reject or accept a model; it seems accretion of
the primary wind by the secondary is inevitable.
In rejecting the $\omega\simeq 90^\circ$ orientation, P09 gave too much
weight to the X-ray minimum period.
\end{enumerate}

Recent observations of the RXTE lightcurve of $\eta$ Car reveal an early recovery
of the X-ray minimum (Corcoran 2009).
While the previous two X-ray minima lasted for
10~weeks, the 2009 X-ray minimum lasted for only $\sim 5$~weeks.
The possibility that the X-ray emission can recover several days earlier
than in previous cycles was mentioned by us before (section 8.3 in Kashi \& Soker 2009a),
as we were considering small ($\sim 10\%$) fluctuations in the primary wind properties.
The early recovery is attributed to a weaker primary wind that allows the
secondary wind to recover earlier.
In the accretion model the very early recovery teaches us that the major
changes in the primary wind are related to the acceleration process (Kashi \& Soker 2009a),
and that the variations are large in the sense that the wind reaches its
terminal speed much closer to the primary.
A quantitative study is the subject of a forthcoming paper.

The implication of the accretion process goes beyond present day $\eta$ Car.
During the 1837-1856 Great Eruption a mass of $\sim 10-20 M_\odot$ was
lost by the primary (smith et al. 2003; Smith 2006; Smith \& Ferland 2007).
If accretion occurs at present periastron passages, when mass loss rate is
lower by more than three orders of magnitudes than that during the Great Eruption,
it must have occurred during the Great Eruption along most,
or even all, of the orbit (Soker 2001, 2007).
The gravitational energy released by the accreted mass could have been the
major extra energy in the Great Eruption, both in extra radiation and wind's
kinetic energy (Soker 2007).
It is possible that major eruptions of luminous blue variables (LBVs) are
related to such accretion events as the primary losses high amounts of mass.

We thank Ehud Behar for helping us with the calculations of the X-ray emission.
We also thank Ross Parkin and Julian Pittard and an anonymous referee for useful comments.
This research was supported by grants from the Israel Science Foundation,
and from Asher Space Research Institute at the Technion.


\begin{references}

\reference{} Abraham, Z., Falceta-Gon\c{c}alves, D., Dominici, T. P.,
Nyman, L.-A, Durouchoux, P., McAuliffe, F., Caproni, A., \&
Jatenco-Pereira, V. 2005, A\&A, 437, 977

\reference{} Akashi, M., Soker, N., \& Behar, E. 2006, ApJ, 644., 451 (A06)

\reference{} Behar, E., Nordon, R., \& Soker, N. 2007, ApJ, 666, L97

\reference{} Corcoran, M. F. 2005, AJ, 129, 2018

\reference{} Corcoran, M. F. 2009, in The RXTE X-ray Lightcurve of Eta Carinae site
\newline
$\texttt{(http://asd.gsfc.nasa.gov/Michael.Corcoran/eta\_car/etacar\_rxte\_lightcurve/index.html)}$

\reference{} Corcoran, M. F., Ishibashi, K., Swank, J. H., \&
 Petre, R., 2001, ApJ, 547, 1034

\reference{} Damineli, A., Conti, P. S., \& Lopes, D. F. 1997, NewA, 2, 107

\reference{} Damineli, A., Hillier, D. J., Corcoran M. F. et al. 2008
      MNRAS, 384, 1649 

\reference{} Davidson, K. 1997 NewA, 2, 387D

\reference{} Davidson, K., Smith, N., Gull, T.R., Ishibashi, K., \&
Hillier, D.J., 2001, AJ, 121, 1569.

\reference{} Dorland, B. N., 2007, PhDT, 9D, ``An Astrometric Analysis
of Eta Carinae's Eruptive History Using HST WF/PC2 and ACS
Observations''

\reference{} Duncan, R. A., \& White, S. M. 2003, MNRAS, 338, 425

\reference{} Eichler, D., \& Usov, V. 1993, ApJ, 402, 271

\reference{} Falceta-Gon\c{c}alves, D., Jatenco-Pereira, V., \&
Abraham, Z. 2005, MNRAS, 357, 895

\reference{} Hamaguchi, K. Corcoran, M. F., Gull, T., Ishibashi, K.,
    Pittard, J. M., Hillier, D. J., Damineli, A., Davidson, K., Nielsen, K.
    E. \& Kober, G. V., 2007, ApJ, 663, 522 (H07)

\reference{} Henley, D. B., Corcoran, M. F., Pittard, J. M., Stevens,
I. R., Hamaguchi, K., \& Gull T. R. 2008, ApJ, 680, 705

\reference{} Hillier, D. J., Gull, T., Nielsen, K., Sonneborn, G.,
Iping, R., Smith, N., Corcoran, M., Damineli, A., Hamann, F. W.,
Martin, J. C., \& Weis, K. 2006, ApJ, 642, 1098  

\reference{} Kashi, A., \& Soker, N. 2007, NewA, 12, 590 

\reference{} Kashi, A., \& Soker, N. 2008, MNRAS, 390, 1751 (KS08) 

\reference{} Kashi, A., \& Soker, N. 2009a, NewA, 14, 11  

\reference{} Kashi, A., \& Soker, N. 2009b, accepted by MNRAS, (arXiv:0808.4132) 

\reference{} Fernandez Lajus, E., Schwartz, M., Salerno, N.; Torres, A., Farina, C.,
 Llinares, C., Calderon, J. P., Bareilles, F., Gamen, R., Niemela, V. S. 2008, A\&A, preprint.

\reference{} Nielsen, K. E., Corcoran, M. F., Gull T. R., Hillier, D.
J., Hamaguchi, K., Ivarsson, S. \& Lindler, D. J. 2007, ApJ, 660, 669

\reference{} Okazaki, A. T., Owocki, S. P., Russell, C. M., P. \& Corcoran,
M. F. 2008,
in Massive Stars as Cosmic Engines, IAU Sym. 250, eds. F. Bresolin, P.A.
Crowther and J. Puls (Cambridge University Press), 133 (arXiv:0803.3977)

\reference{}  Parkin, E. R., Pittard, J. M., Corcoran, M. F., Hamaguchi, K.,
     \& Stevens I. R. 2009, MNRAS in press (arXiv:0901.0862) (P09)

\reference{} Pittard, J. M., Corcoran, M. F. 2002, A\&A 383 636P

\reference{} Pittard, J. M., Stevens, I. R., Corcoran, M. F., \& Ishibashi,
K. 1998, MNRAS, 299, L5

\reference{} Sarazin, C. L., \& Bahcall, J. N.  1977, ApJS, 34, 451

\reference{} Smith, N. 2002, MNRAS, 337, 1252

\reference{} Smith, N. 2006, ApJ, 644, 1151

\reference{} Smith, N., \& Ferland, G. J.  2007, ApJ, 655, 911

\reference{} Smith, N., Gehrz, R., D., Hinz, P. M., Hoffmann, W. F., Hora,
J. L., Mamajek, E. E.,
    \& Meyer, M. R. 2003, AJ, 125, 1458

\reference{} Smith, N., Morse, J. A., Collins, N. R., \& Gull, T. R.
      2004, ApJ, 610, L105

\reference{} Soker, N. 2001, MNRAS, 325, 584

\reference{} Soker, N. 2005, ApJ, 635, 540  

\reference{} Soker, N. 2007, ApJ, 661, 490  

\reference{} Soker, N., \& Behar, E. 2006, ApJ, 652, 1563

\reference{} van Genderen A. M., Sterken, C., Allen, W. H., \& Walker,
W. S. G. 2006, JAD, 12, 3

\reference{} Whitelock, P. A., Feast, M. W., Marang, F., \& Breedt, E.
2004, MNRAS, 352, 447

\end{references}
\end{document}